\documentclass[paper,notoc,nohyper]{JHEP}
\usepackage{epsfig}
\renewcommand\O[1]{{\cal O}(#1)}
\def\S{S_{\epsilon}}

\def\bzero{\beta_0}

\def\Bfin{{\rm Box^6} (u, t) }
\def\Bubl{{\rm Bub}(s)}

\def\lib{{\rm Li}_2}
\def\lic{{\rm Li}_3}
\def\lid{{\rm Li}_4}

\def\lnx{L_x}
\def\lny{L_y}

\def\Libx{{\rm Li}_2(x)}

\def\Licx{{\rm Li}_3(x)}

\def\Lidx{{\rm Li}_4(x)}

\def\CA{C_A}
\def\CF{C_F}

\def\A{{\cal A}}
\def\B{{\cal B}}
\renewcommand\O[1]{{\cal O}(#1)}
\def\as{\ensuremath{\alpha_{s}}}

\def\Re{\mathop{\rm Re}}

\def\t3ou{\,\frac{t^3}{us^2}}
\def\u3ot{\,\frac{u^3}{ts^2}}

\def\NF{\,N_F}

\def\beq{\begin{equation}}
\def\eeq{\end{equation}}

\def\beqn{\begin{eqnarray}}
\def\eeqn{\end{eqnarray}}

\def\lq{\left[}
\def\rq{\right]}

\def\({\left(}
\def\){\right)}

\def\ket#1{|{#1}\rangle}

\def\braket#1#2{\langle #1 |#2 \rangle}
\def\cm{{\cal M}}
\def\cmb{\overline{{\cal M}}}
\def\t#1#2#3{t^#1_{#2\,#3}}

\def\a0{\alpha_0}

\def\ab0{\bar{\alpha}_0}

\def\fs{\(-\frac{\mu^2}{s}\)^\ep }
\def\ft{\(-\frac{\mu^2}{t}\)^\ep }
\def\fu{\(-\frac{\mu^2}{u}\)^\ep }

\def \ep{\epsilon}
\def\ord#1{{\cal O}\(#1\)}
\def\MSbar{$\overline{{\rm MS}}$}

\def\irt{{\cal IR}_{t}}
\def\irnt{{\cal IR}_{nt}}

\def\irbb{{\left( \overline{{\cal IR}}_{nt} - \overline{{\cal F}}_{r} 
-\overline{{\cal F}}_{g} \right)}^{\dagger}}
\def\treef{\left( {\cal IR}_{t} + {\cal F}_{r} + {\cal F}_{g} \right)}
\def\treefb{\left( \overline{{\cal IR}}_{t} + {\overline{{\cal F}}_{r}} +
{\overline{{\cal F}}_{g}} \right)^{\dagger}}
\def\absq#1{ {\left | #1 \right |}^2}

\newcommand{\BUB}[1]{
\mbox{\parbox{3cm}{
\begin{picture}(3.5,1.4)
\thicklines
\put(0.5,0.7){\line(1,0){0.5}}
\put(2.0,0.7){\line(1,0){0.5}}
\put(1.5,0.7){\circle{1}}
\put(2.65,0.7){\makebox(0,0)[l]{$(#1)$}}
\end{picture}
}}
\hfill}

\newcommand{\BOX}[2]{
\mbox{\parbox{3cm}{
\begin{picture}(3.5,1.4)
\thicklines
\put(0.5,0.2){\line(1,0){2.4}}
\put(0.5,1.2){\line(1,0){2.4}}
\put(1.2,0.2){\line(0,1){1}}
\put(2.2,0.2){\line(0,1){1}}
\put(3.05,0.7){\makebox(0,0)[l]{$(#1,#2)$}}
\end{picture}
}} 
\hfill}
\title{\boldmath One-loop QCD corrections to quark scattering at 
NNLO\footnote{Work supported in part by the UK Particle Physics and Astronomy 
Research Council and by the EU Fourth Framework Programme
`Training and Mobility of Researchers', Network `Quantum Chromodynamics
and the Deep Structure of Elementary Particles',
contract FMRX-CT98-0194 (DG 12 - MIHT).
C.A. acknowledges
the financial support of the Greek government and
M.E.T. acknowledges financial support
from CONACyT and the CVCP. We thank
the British Council and German Academic Exchange Service for support
under ARC project 1050.
}}
\author{
C.~Anastasiou$^a$,
E.~W.~N.~Glover$^a$,
C.~Oleari$^b$ and M.~E.~Tejeda-Yeomans$^a$\\
$^a$Department of Physics, 
University of Durham, 
Durham DH1 3LE, 
England\\[1mm]
$^b$Department of Physics, 
University of Wisconsin,
1150 University Avenue\\
Madison WI 53706, 
U.S.A.\\[1mm]
E-mail: \email{Ch.Anastasiou@durham.ac.uk}, \email{E.W.N.Glover@durham.ac.uk},
\email{Oleari@pheno.physics.wisc.edu}, 
\email{M.E.Tejeda-Yeomans@durham.ac.uk}}
\abstract{
We present the $\O{\as^4}$ virtual QCD corrections to unlike-quark $q \bar q \to 
q^\prime \bar{q}^\prime$ and like-quark scattering $q \bar{q} \to 
q \bar{q}$  due to the interference of one-loop amplitudes 
with one-loop amplitudes.
The structure of the infrared divergences agrees with that predicted 
by Catani.  The results are expressed in an analytic form so that the relevant
expressions for crossed scattering processes can be obtained in a 
straightforward manner.   The one-loop contributions 
presented here, together  with 
the interference of tree with two-loop amplitudes given recently, 
complete the $2\to 2$  virtual QCD corrections at NNLO for the massless
quark scattering processes.} 
\preprint{DTP/00/76, {IPPP/00/11}, {MADPH-00-1206}, {hep-ph/0012007}}

\begin{document} \date 

\section{Introduction} 
\label{sec:intro}

The increasing precision of high energy scattering experiments demonstrates 
the need for very precise theoretical calculations.   The accuracy
of existing next-to-leading-order (NLO)  predictions may be
improved  by including the next highest term in the  perturbation  series. 
Such next-to-next-to-leading-order (NNLO) estimates  will improve the
theoretical precision in two ways. First, the renormalisation  scale
dependence will  decrease (from about $10\%$ for NLO predictions  for central
production  of a jet with transverse energy  $E_T \sim 100$~GeV at the
Tevatron to approximately $1-2\%$ at NNLO). Second, better matching of the
parton level  theoretical  and hadron level experimental jet algorithms will be
possible since at NNLO, three partons can combine to form the jet rather 
than two.

Recent progress towards  two-loop 
%~\cite{planarA, 
%nonplanarA, planarB, nonplanarB, AGO2, AGO3, bastei3, bastei2}  
integrals~\cite{planarA}--\cite{bastei2}  
has turned  
the calculation  of NNLO virtual corrections  for massless $2 \to 2$ processes 
into a viable task.  Bern, Dixon and Kosower~\cite{bdk} were the first to
address such scattering processes and provided analytic expressions for the
maximal-helicity-violating two-loop amplitude for $gg \to gg$.   More recently,
Bern, Dixon and Ghinculov~\cite{BDG}  completed the two-loop calculation of
physical $2 \to 2$ scattering amplitudes for the QED  processes $e^+e^- \to
\mu^+\mu^-$ and $e^+e^- \to e^-e^+$. 

Subsequently, we have derived analytical expressions  for the 
${\cal O}(\alpha_s^4)$ 
two-loop contribution to unlike quark scattering  $q \bar q \to 
q^\prime \bar{q}^\prime$~\cite{qqQQ}, and like quark scattering  $q \bar q \to 
q\bar{q}$~\cite{qqqq}, as well as the crossed and time reversed  processes, in
the limit where the quark mass can be neglected. To
complete the calculation of the NNLO virtual corrections,  the interference
terms of one-loop  amplitudes  with one-loop amplitudes  need to be included. 
It is the purpose of this paper to present analytical expressions  for these
contributions using conventional dimensional regularisation (space-time
dimension $D=4-2\ep$), renormalised with the \MSbar\ scheme.   

Our paper is organised as follows. We establish our notation in 
Section~\ref{sec:notation}  and briefly describe our method in
Section~\ref{sec:method}.  In Sections~\ref{sec:unlike} and~\ref{sec:like}, we
provide  analytic expressions of the one-loop contributions at NNLO for the 
unlike and like-quark scattering respectively, obtained  by direct  evaluation
of the Feynman diagrams.  Our results are expressed in terms of two master
integrals, the one-loop  box graph in  $6-2\ep$ dimensions and the one-loop
bubble graph  in $4-2\ep$ dimensions. A clear separation of the infrared
poles   is apparent  by inspection  of the results. We find complete  agreement
between the infrared structure obtained by our explicit calculation and  that
anticipated by  Catani~\cite{catani}. Analytic expansions in
$\ep$ for all  kinematic regions are straightforward to derive by
inserting the expansions of the master  integrals in the appropriate region. 
In  Section~\ref{sec:conclusions} we summarize our results.

\section{Notation}
\label{sec:notation}
We follow the notation of Refs.~\cite{qqQQ,qqqq} as closely as possible and
consider the  unlike-quark scattering process
\begin{equation}
\label{eq:proc1}
q (p_1) + \bar q (p_2)  + q^\prime (p_3) + \bar{q}^\prime(p_4) \to 0,
\end{equation}
and the like-quark scattering process 
\begin{equation}
\label{eq:proc2}
q (p_1) + \bar q (p_2)  + q (p_3) + \bar{q}(p_4) \to 0,
\end{equation}
where  particles are incoming and carry  light-like  momenta (shown 
in parentheses). Their total momentum is conserved, satisfying
$$
p_1^\mu+p_2^\mu+p_3^\mu+p_4^\mu = 0,
$$
and the associated Mandelstam variables are given by
\begin{equation}
s = (p_1+p_2)^2, \qquad t = (p_2+p_3)^2, \qquad u = (p_1+p_3)^2.
\end{equation}
We use conventional dimensional regularisation  and treat the external quark
states in $D$ space-time dimensions and  renormalise the ultraviolet
divergences in the \MSbar\ scheme.  The bare coupling
$\a0$ is related to the running coupling  $\as \equiv \alpha_s(\mu^2)$,  
at renormalisation
scale $\mu$, by 
\beq
\label{eq:alpha}
\a0 \,  \S = \as \,  \lq 1 - \frac{\beta_0}{\ep}  
\, \left(\frac{\as}{2\pi}\right) + \( \frac{\beta_0^2}{\ep^2} - 
\frac{\beta_1}{2\ep} \)  \, 
\left(\frac{\as}{2\pi}\right)^2
+\O{\as^3} \rq,
\eeq
where 
\beq
\S = (4 \pi)^\ep e^{-\ep \gamma},  \quad\quad \gamma=0.5772\ldots=
{\rm Euler\ constant },
\eeq
is the typical phase-space volume factor in $D=4-2\ep$ dimensions.
As usual, the first two coefficients of the QCD beta function, $\beta_0$ and 
$\beta_1$  for $\NF$ (massless) quark flavours
are
\beq
\label{betas}
\beta_0 = \frac{11 \CA - 4 T_R \NF}{6} \;\;, \;\; \;\;\;\;
\beta_1 = \frac{17 \CA^2 - 10 \CA T_R \NF - 6 \CF T_R \NF}{6} \;\;.
\eeq
where $N$ is the number of colours, and 
\beq
\CF = \frac{N^2-1}{2N}, \qquad \CA = N, \qquad T_R = \frac{1}{2}.
\eeq
The renormalised amplitude for the unlike-quark  process is given by 
\beq
\ket{\cm}_{unlike}= 4\pi \as \lq \ket{\cm^{(0)}} + \left(\frac{\as}{2\pi}\right) 
\,\ket{\cm^{(1)}}
+ \left(\frac{\as}{2\pi}\right)^2\,
\ket{ \cm^{(2)}}+ \ord{\as^3} \rq,
\eeq
with $\ket{\cm^{(i)}}$ representing the $i$-loop amplitude in colour-space. 
For the  like-quarks we have the related expression
\beqn
\ket{\cm}_{like}&=& 4\pi \as \Biggl [ \left(\ket{\cm^{(0)}} 
-\ket{\cmb^{(0)}}\right)
+ \left(\frac{\as}{2\pi}\right) \left(\,\ket{\cm^{(1)}}-\ket{\cmb^{(1)}}\right)
\nonumber \\
&& \hspace{4cm}+ \left(\frac{\as}{2\pi}\right)^2\,
\left(\ket{ \cm^{(2)}}-\ket{ \cmb^{(2)}}\right) + \O{\as^3} \Biggr ].
\eeqn
Here $\ket{\cmb^{(i)}}$ describes the $t$-channel graphs 
which can be  obtained  from the $s$-channel diagrams by exchanging the 
roles of particles 2 and 4
\beq
\ket{\cmb^{(i)}} = \ket{\cm^{(i)}} ( 2 \leftrightarrow 4).
\eeq
Both $\ket{ \cm^{(i)}}$  and $\ket{ \cmb^{(i)}}$ are renormalisation 
scale  and renormalisation scheme dependent.

In squaring the amplitudes and summing over colours and spins we find two 
types of terms,
\begin{itemize}
\item the self-interference of the graphs in a single channel,  
described by the function $\A(s, t, u)$ for the $s$-channel and $\A(t,s,u)$ for
the $t$-channel, and
\item  the interference of the $s$-channel graphs with the $t$-channel 
graphs, described by the function $\B(s, t, u)$.
\end{itemize}
Thus, for distinct quark scattering we have 
\begin{equation}
\braket{\cm}{\cm}_{unlike} = 
\sum |{\cal M}({q + \bar{q} \to  \bar{q}^\prime + q^\prime })|^2
=\A(s,t,u),
\end{equation}
while for  identical quarks
\beqn
\braket{\cm}{\cm}_{like} &=& \sum |{\cal M}({q + \bar{q} \to  \bar{q} + q })|^2
\nonumber \\  
&=& \A(s,t,u) + \A(t,s,u) + \B(s,t,u).
\eeqn
Similarly, for the crossed and time-reversed  processes we obtain
\begin{eqnarray}
%\sum |{\cal M}({q + \bar{q} \to  \bar{q}^\prime + q^\prime } )|^2 
%&=&\A(s,t,u) \\
\sum |{\cal M}({q + q^\prime \to  q + q^\prime})|^2 
&=& \A(u,t,s) \\
\sum |{\cal M}({q + \bar{q}^\prime \to  q + \bar{q}^\prime})|^2 
&=& \A(t,s,u) \\
\sum |{\cal M}({\bar{q} + \bar{q}^\prime \to  \bar{q} +\bar{q}^\prime})|^2 
&=& \A(u,t,s) \\
%\sum |{\cal M}({q +\, \bar{q} \to  \bar{q} +\, q })|^2 
%&=& \A(s,t,u) + \A(t,s,u) + \B(s,t,u) \\
\sum |{\cal M}({q +\, q \to  q +\, q })|^2 
&=& \A(u,t,s) + \A(t,u,s) + \B(u,t,s).
\end{eqnarray} 

The function $\A$ can be expanded perturbatively to yield
\beqn
\A(s,t,u) &=& 16\pi^2\as^2 \left[
\A^4(s,t,u)+\left(\frac{\as}{2\pi}\right) \A^6(s,t,u)
+\left(\frac{\as}{2\pi}\right)^2 \A^8(s,t,u) +
\O{\as^{3}}\right],\nonumber \\  
\eeqn
where
\beqn
\A^4(s,t,u) &=& \braket{\cm^{(0)}}{\cm^{(0)}} \equiv 2(N^2-1) 
\left(\frac{t^2+u^2}{s^2} - \epsilon
\right),\\
\A^6(s,t,u) &=& \left(
\braket{\cm^{(0)}}{\cm^{(1)}}+\braket{\cm^{(1)}}{\cm^{(0)}}\right),\\
\A^8(s,t,u) &=& \left( \braket{\cm^{(1)}}{\cm^{(1)}} +
\braket{\cm^{(0)}}{\cm^{(2)}}+\braket{\cm^{(2)}}{\cm^{(0)}}\right). 
\eeqn
In the same manner
\beqn
\B(s,t,u) &=& 16\pi^2\as^2 \left[
 \B^4(s,t,u)+\left(\frac{\as}{2\pi}\right) \B^6(s,t,u)
 +\left(\frac{\as}{2\pi}\right)^2 \B^8(s,t,u) +
\O{\as^{3}}\right],\nonumber \\
\eeqn
where, in terms of the amplitudes, we have
\beqn
\B^4(s,t,u) &=& -\left(\braket{\cmb^{(0)}}{\cm^{(0)}}
 +\braket{\cm^{(0)}}{\cmb^{(0)}}\right) \nonumber \\
&\equiv& -4\left(\frac{N^2-1}{N}\right) (1-\epsilon)
\left(\frac{u^2}{st} + \epsilon\right),\\
\B^6(s,t,u) &=& -\left(\braket{\cmb^{(1)}}{\cm^{(0)}}
+\braket{\cm^{(0)}}{\cmb^{(1)}}
+\braket{\cmb^{(0)}}{\cm^{(1)}} 
+\braket{\cm^{(1)}}{\cmb^{(0)}}\right)\nonumber \\
&&\\
\B^8(s,t,u) &=& -\left(
\braket{\cmb^{(1)}}{\cm^{(1)}}
+\braket{\cm^{(1)}}{\cmb^{(1)}}\right.\nonumber \\
&& \hspace{0.5cm}
\left.
+\braket{\cmb^{(0)}}{\cm^{(2)}}
+\braket{\cm^{(2)}}{\cmb^{(0)}}
+\braket{\cm^{(0)}}{\cmb^{(2)}}
+\braket{\cmb^{(2)}}{\cm^{(0)}}
\right).\nonumber \\
\eeqn
Expressions  for $\A^6$ and $\B^6$, valid in 
dimensional regularisation,  are given in Ref.~\cite{ES}.  
Analytical formulae for the two-loop contribution to $\A^8$
\[
\braket{\cm^{(0)}}{\cm^{(2)}}+\braket{\cm^{(2)}}{\cm^{(0)}}  
\]
are given in Ref.~\cite{qqQQ}, with analogous expressions
for the two-loop contribution to $\B^8$
\[
\braket{\cmb^{(0)}}{\cm^{(2)}}
+\braket{\cm^{(2)}}{\cmb^{(0)}}
+\braket{\cm^{(0)}}{\cmb^{(2)}}
+\braket{\cmb^{(2)}}{\cm^{(0)}}
\]
given in Ref.~\cite{qqqq}. 

We now concentrate on the contributions to 
both $\A^8$ and $\B^8$ due to the interference of one-loop amplitudes with 
one-loop amplitudes, namely
\begin{equation}
\A^{8 \, (1 \times 1)}(s, t, u) = \braket{\cm^{(1)}}{ \cm^{(1)}},
\end{equation}
and
\begin{equation}
\B^{8 \, (1 \times 1)}(s, t, u) = 
-\left(
\braket{\cmb^{(1)}}{ \cm^{(1)}}+\braket{\cm^{(1)}}{ \cmb^{(1)}}
\right).
\end{equation}
Even though they are somewhat simpler to evaluate than the two loop graphs,
they form a vital part of the NNLO virtual corrections and we present them here
for completeness. One-loop helicity amplitudes for the $2 \rightarrow 2$ quark
scattering processes   were given in Ref.~\cite{adrian} as truncated expansions
in $\ep$ including  their finite part. However, this is only sufficient to
obtain the pole  structure of $\A^{8 \, (1 \times 1)}$ and $\B^{8 \, (1 \times
1)}$ up to $1/\ep^2$.  To determine the $1/\ep$ and finite parts  requires
knowledge of the one-loop  amplitude through to ${\cal O}(\ep^2)$.    

\section{Method}
\label{sec:method}
We organise our calculation  as follows. 
First the one-loop Feynman diagrams are generated using {\tt QGRAF} 
\cite{QGRAF}. The emerging tensor integrals are associated with scalar 
integrals in higher dimension and with higher powers of 
propagators~\cite{AGO3, Tar}. Systematic application of 
the integration-by-parts 
identities~\cite{IBP} is sufficient to reduce these higher-dimension, 
higher-power integrals to master integrals in $D=4-2 \ep$. We can express all
the integrals of the one-loop amplitudes in terms of just two master 
integrals, the scalar bubble graph
$$
{\rm Bub}(s) = \BUB{s},
$$
and the one-loop scalar box graph
$$
{\rm Box}(s,t) = \BOX{s}{t}.
$$ 
The above choice of master integrals is not unique.  We prefer to replace the
one-loop box in $D=4-2\ep$ by the finite one-loop box in $D=6-2 \ep$, ${\rm
Box}^6$. This leads to a natural  separation of  the infrared poles and the
finite part of the one-loop amplitudes. As a last step we multiply together the
one-loop amplitudes and perform the colour and Dirac traces.

\section{Results}
\label{sec:results}
In this section we give explicit formulae for both $\A^{8 \, (1 \times 1)}(s,
t, u)$ and $\B^{8 \, (1 \times 1)}(s, t, u) $, in terms of the finite  ${\rm
Box}^6$ and the $1/\ep$  divergent ${\rm Bub}$ master integrals. 
The finite parts depend only on ${\rm Box}^6$ and differences of the 
${\rm Bub}$ master integrals. 

\subsection{Unlike quarks}
\label{sec:unlike}
In the unlike-quark case we obtain,
\begin{eqnarray}
\A^{8 \, (1 \times 1)}(s, t, u) &=& 
\left[ \absq{\irt+{\cal F}_{r}+{\cal F}_{g}} + \left(N^2-1\right) 
\absq{\irnt} \right] \braket{\cm_0}{\cm_0} \nonumber \\   
&+& 2 \Re \left[ \left(\irt+{\cal F}_{r}+{\cal F}_{g}\right)^{\dagger} {\cal 
F}_1
+ \left( N^2-1\right) \irnt^{\dagger} {\cal F}_2
\right] \nonumber \\
&+& \left(N^2-1 \right) \Bigg[ \frac{N^4-3N^2+3}{N^2} {\cal F}_3(s, t, u)
+\frac{N^2+3}{N^2} {\cal F}_3(s, u, t) \nonumber \\
&& \hspace{2cm} +\frac{N^2-3}{N^2}  \left[ {\cal F}_4(s, t, u) 
   + {\cal F}_4(s, u, t) \right] 
\Bigg],
\label{eq:unlike}
\end{eqnarray}
where the infrared poles present in the one-loop amplitude 
proportional to the tree-level matrix elements are given by 
\begin{eqnarray}
\irt &=& \frac{2}{\ep (2+\ep)} \left[
\frac{1}{N} {\rm Bub}(s)
-\frac{2}{N} {\rm Bub}(u)
-\frac{(N^2-2)}{N} {\rm Bub}(t)
\right],\\
\irnt&=& \frac{2}{\ep (2+\ep)} \left[
\frac{1}{N}{\rm Bub}(u) - \frac{1}{N} {\rm  Bub}(t)
\right],
\end{eqnarray}
which diverge as $1/\ep^2$ and  $1/\ep$ respectively.
Both
\begin{equation} 
{\cal F}_{r} =\bzero 
\left(-\frac{1}{\ep} +\frac{3(1-\ep)}{3-2\ep} ~\Bubl \right),
\end{equation}
and
\begin{equation}
{\cal F}_{g}=\frac{\ep \left[N^2 (11+2\ep)+9-4\ep^2 \right]}
{2 (2+\ep) (3-2\ep)N} ~\Bubl,
\end{equation}
are finite terms multiplying the tree-level matrix elements.
The functions 
\begin{equation}
{\cal F}_1 = \frac{N^2-1}{2 N} \left[(N^2-2) f(s, t, u) +2 f(s, u, t)\right], 
\end{equation}
and
\begin{equation}
{\cal F}_2 = \frac{N^2-1}{2 N} \Big[ f(s, t, u) -f(s, u, t)\Big]
\end{equation}
are finite and multiplied by the infrared poles of the conjugated
one-loop  amplitude, with
\begin{eqnarray}
f(s, t, u)&=& \left[ \frac{3s^2+7u^2+9t^2}{s^2} -4 \frac{u^2+t^2+2 s^2}{(2+\ep) 
s^2} +\ep \frac{5 u +7 t}{s}\right] \Big[ {\rm Bub}(t)-{\rm Bub}(s)\Big]
\nonumber \\
&& +u (1-2 \ep) \frac{6 t^2 +2 u^2 -3 \ep s^2}{s^2} ~{\rm Box}^6(s, t).
\end{eqnarray}
Finally the square of the finite part of the one-loop amplitude is fixed by
the finite functions ${\cal F}_3$ and ${\cal F}_4$,
\begin{eqnarray}
{\cal F}_3(s, t, u)&=& \absq{{\rm Box}^6(s, t)} 
\left[ \frac{t^4+6t^2 u^2+u^4}{2 s^2} \right]
\nonumber \\
&+& 2 \Re \left\{ \Big[{\rm Bub}(t)-{\rm Bub}(s) \Big]^{\dagger}
 ~{\rm Box}^6(s, t) \right\} 
\left[ 
\frac{2 u^3-t u^2 +8 t^2 u-t^3}{2 s^2}
\right]
\nonumber \\
&+&  \absq{ {\rm Bub}(t) -{\rm Bub}(s) } 
\left[ \frac{5 t^2-2 t u +2 u^2}{s^2}\right] +{\cal O}(\ep), 
\end{eqnarray}
and 
\begin{eqnarray}
{\cal F}_4(s, t, u)&=& 2 \Re \left\{ 
{{\rm Box}^6}^{\dagger}(s, t) {\rm Box}^6(s, u) \right\} 
\left[
\frac{t u (t^2+u^2)}{s^2}
\right]
\nonumber \\
&&\hspace{-1.5cm} +2 \Re \left\{ 
\left[ {\rm Bub}(u) - {\rm Bub}(s)\right]^{\dagger} ~{\rm Box}^6(s, t)
\right\} 
\left[
\frac{u (7 t^2-2tu+3u^2)}{2 s^2} 
\right]
\nonumber \\
&& \hspace{-1.5cm} +2 \Re \left\{ 
\left[ {\rm Bub}(u) - {\rm Bub}(s)\right]^{\dagger}
\left[ {\rm Bub}(t) - {\rm Bub}(s)\right]
\right\} 
\left[
\frac{3 (t^2-tu+u^2)}{2 s^2}
\right] + {\cal O}(\ep).
\nonumber \\
\end{eqnarray}
In the latter expressions, we have discarded contributions of ${\cal O}(\ep)$.
 
After explicit series expansion in $\ep$, the infrared singular terms $\irt$ and 
$\irnt$ reproduce the pole structure obtained by expanding 
\beqn
{\cal IR}_{t, C} &=&  \frac{e^{\ep \gamma}}{\Gamma(1-\ep)}
\left( 
\frac{1}{\ep^2} +\frac{3}{2 \ep}
\right) 
\Bigg[
\frac{1}{N} \fs
-\frac{2}{N} \fu
-\frac{(N^2-2)}{N} \ft
\Bigg], \nonumber \\ \\
{\cal IR}_{nt, C} &=& 
\frac{e^{\ep \gamma}}{\Gamma(1-\ep)}
\left(
\frac{1}{\ep^2} +\frac{3}{2 \ep}
\right)  \Bigg[  \frac{1}{N}\fu - \frac{1}{N} \ft \Bigg],
\eeqn
which is the singular structure obtained by straightforward application of
the formalism of~\cite{catani}. To rewrite
Eq.~(\ref{eq:unlike}) directly in terms of ${\cal IR}_{t, C}$ and
${\cal IR}_{nt, C}$ rather than $\irt$ and $\irnt$ requires the finite 
difference to be evaluated through to  ${\cal O}(\ep^2)$.

Equation~(\ref{eq:unlike}) is valid in all kinematic 
regions.  Series expansions   in $\ep$  in a  particular region 
can be easily obtained by inserting the appropriate expansions of the 
master integrals. In this equation, the finite functions are multiplied by poles
in $\ep$, so they must be expanded through to  ${\cal O}(\ep^2)$.
In the physical region $ u < 0$, $t < 0$, 
$\Bfin$ has no imaginary part and is given by~\cite{BDG}
\begin{eqnarray}
\Bfin &=& \frac{ e^{\ep\gamma}
\Gamma  \left(  1+\epsilon \right)  \Gamma  
\left( 1-\epsilon \right) ^2 
 }{ 2s\Gamma  \left( 2-2 \epsilon  \right)} \fs
 \times 
  \Biggl [
 \frac{1}{2}((\lnx-\lny)^2+\pi^2 )\nonumber \\
&& 
 +2\ep \left(
 \Licx-\lnx\Libx-\frac{1}{3}\lnx^3-\frac{\pi^2}{2}\lnx \right)
\nonumber \\
&& 
-2\ep^2\biggl(
\Lidx+\lny\Licx-\frac{1}{2}\lnx^2\Libx-\frac{1}{8}\lnx^4-\frac{1}{6}\lnx^3\lny+\frac{1}{4}\lnx^2\lny^2\nonumber
\\
&&-\frac{\pi^2}{4}\lnx^2-\frac{\pi^2}{3}\lnx\lny-\frac{\pi^4}{45}\biggr)
+ ( u \leftrightarrow t) \Biggr ] + \O{\ep^3},
\end{eqnarray}
where $x = -t/s$, $\lnx=\log(x)$ and $\lny=\log(1-x)$ and the 
polylogarithms ${\rm Li}_n(z)$ are defined by,
\begin{eqnarray}
 {\rm Li}_n(z) &=& \int_0^z \frac{dt}{t} {\rm Li}_{n-1}(t) \qquad {\rm ~for~}
 n=2,3,4\\
 {\rm Li}_2(z) &=& -\int_0^z \frac{dt}{t} \log(1-t).
\end{eqnarray} 
Analytic continuation to other kinematic regions is obtained 
using the inversion formulae for the arguments of 
the polylogarithms (see for example~\cite{AGO3}) when $x > 1$,
\begin{eqnarray}
\lib(x + i0) &=& -\lib\left(\frac{1}{x}\right) -\frac{1}{2}
 \log^2 (x) +\frac{\pi^2}{3} + i \pi \log (x) \nonumber \\
\lic(x + i0) &=& \lic\left(\frac{1}{x}\right) -\frac{1}{6}\log^3(x) +
\frac{\pi^2}{3}  \log(x) +  \frac{i\pi}{2} \log^2(x) \nonumber \\ 
%%%%%%%%%%%%%
\lid(x + i0)  &=& - \lid\left(\frac{1}{x}\right)
         -  \frac{1}{24}\log^4(x)  + \frac{\pi^2}{6} \log^2(x)
        + \frac{\pi^4}{45} +
        \frac{i\pi}{6}\log^3(x) 
\end{eqnarray}
Finally, the one-loop bubble integral in $D=4-2 \epsilon$ dimensions 
is given by,
\begin{equation} 
 \Bubl =\frac{ e^{\ep\gamma}\Gamma  \left(  1+\epsilon \right)  \Gamma  \left( 
1-\epsilon \right) ^2 
 }{ \Gamma  \left( 2-2 \epsilon  \right)  \epsilon   } \fs,
\end{equation}
and can be easily expanded in $\ep$ in all kinematic regions.
Note that the complex conjugate of the master integrals is also required
in Eq.~(\ref{eq:unlike}).

\subsection{Like quarks}
\label{sec:like}
For the like-quark contribution we find a similar expression,
\begin{eqnarray}
\B^{8 \, (1 \times 1)}(s, t, u)&=& \nonumber \\
&& \hspace{-2.7cm} -2 \Re \Bigg\{ 
\treefb \treef \braket{\cmb_0}{\cm_0}
\nonumber \\
&& \hspace{-1.5cm}
+(N^2-1)\irbb \irnt \, \braket{\cmb_0}{\cm_0}
\nonumber \\
&& \hspace{-1.5cm}
+\left[
\treefb {\cal F}^{'}_1 + (N^2-1) \irbb {\cal F}^{'}_2 + \left( s \leftrightarrow 
t \right) \right]
\nonumber \\
&& \hspace{-1.5cm}
+\frac{N^2-1}{N} \Bigg[
-\frac{N^4-N^2-1}{2 N^2} \, f^{\dagger}_3(s, t, u) f_3(t, s, u)
\nonumber \\
&& \hspace{0.3cm} +\frac{N^4-2N^2-1}{2 N^2} \left[f^{\dagger}_3(s, t, u) f_4(t, 
s, u)+ (s \leftrightarrow t) \right]
\nonumber \\
&& \hspace{0.3cm}
+\frac{3 N^2+1}{2 N^2}  \,f^{\dagger}_4(s, t, u) f_4(t, s, u)
\Bigg]
\Bigg\}.
\label{eq:like}
\end{eqnarray}
The infrared singular functions are given by
\begin{eqnarray}
\overline{{\cal IR}}_{t}&=&\frac{2}{\ep (2+\ep)} \Bigg[
\frac{1}{N} {\rm Bub}(s)
+\frac{1}{N} {\rm Bub}(t)
-\frac{(N^2+1)}{N} {\rm Bub}(u)
\Bigg] ,\\
\overline{{\cal IR}}_{nt}&=&\frac{2}{\ep (2+\ep)} \Bigg[
\frac{N^2-1}{N} {\rm Bub}(s)
-\frac{1}{N} {\rm Bub}(t)
+\frac{1}{N} {\rm Bub}(u)
\Bigg],
\end{eqnarray}
which diverge as $1/\ep^2$. The finite renormalisation term is
\begin{equation}
\overline{{\cal F}}_r=\bzero 
\left(-\frac{1}{\ep} +\frac{3(1-\ep)}{3-2\ep} {\rm Bub}(t) \right) ,
\end{equation}
while the remaining finite contribution multiplying tree-level is given by
\begin{equation}
\overline{{\cal F}}_g=
\frac{\ep \left(N^2 (11+2\ep)+9-4\ep^2 \right)}{2 (2+\ep) (3-2\ep)N} 
{\rm Bub}(t).
\end{equation}
Once again, the finite part of the crossed one loop amplitude 
multiplying the infrared
divergent terms of the one loop amplitude generates finite functions
\begin{equation}
{\cal F}^{'}_1 = \frac{N^2-1}{2 N^2} \left[ 
\left(N^2-2\right) f_1(s, t, u) + 2 f_2(s, t, u)
\right],
\end{equation}
and 
\begin{equation}
{\cal F}^{'}_2 = \frac{N^2-1}{2 N^2} \left[ 
f_1(s, t, u) - f_2(s, t, u)
\right],
\end{equation}
where 
\begin{eqnarray}
f_1(s, t, u)&=&\frac{2 u}{s t} (1-2 \ep) \left[
t^2+u^2-2 \ep (t^2+s^2) +\ep^2 s^2
\right] {\rm Box}^6(s, t)
\nonumber \\
&& +\frac{2 }{s t (2+ \ep)} \left[
2 u(2 u-t)+ \ep (u^2-t u -4t^2) +\ep^2 (t s -4 u^2) \right.
\nonumber \\
&& \hspace{1.9cm} \left.
+ \ep^3 u s + \ep^4 ts
\right] \left[ {\rm Bub}(t) -{\rm Bub}(s) \right],
\end{eqnarray}
and
\begin{eqnarray}
f_2(s, t, u)&=& \frac{2}{s} (1-2 \ep) \left[
2 u^2 -\ep (t^2+s^2+u^2) +3 s^2 \ep^2+s^2 \ep^3
\right] {\rm Box}^6(s, u)
\nonumber \\
&&  \frac{2}{t s (2+ \ep)}\left[
6 u^2 -2 t^2 \ep -\ep^2 (2 t^2 +5 u^2 + 3 t u)-\ep^3 s^2 \right.
\nonumber \\
&& \hspace{1.9cm}
\left.
 +\ep^4 t s   
\right] \left[ 
{\rm Bub}(u) -{\rm Bub}(s)
\right].
\end{eqnarray}
Finally the square of the finite part of the one-loop amplitude is controlled
 by
the finite functions $f_3$ and $f_4$
\begin{equation}
f_3(s, t, u) = \frac{1}{s} \left\{(s^2+u^2) {\rm Box}^6(s, t) + 
(2 u-s) \Big[ {\rm Bub}(s) -{\rm Bub}(t) \Big]
\right\}+{\cal O}(\ep),
\end{equation}
and
\begin{equation}
f_4(s, t, u) = \frac{u}{s} \left\{2 s {\rm Box}^6(t, u) + 
3 \Big[ {\rm Bub}(u) -{\rm Bub}(t) \Big]
\right\}+{\cal O}(\ep).
\end{equation}
Again, the infrared singular structure obtained by explicit expansion of 
$\overline{{\cal IR}}_{t}$ and $\overline{{\cal IR}}_{nt}$ as series 
in
$\ep$, agrees with that obtained using Catani's formalism~\cite{catani}
\begin{equation}
\overline{{\cal IR}}_{t,C}=
\frac{e^{\ep \gamma}}{\Gamma(1-\ep)}
\left( 
\frac{1}{\ep^2} +\frac{3}{2 \ep}
\right) 
\Bigg\{
\frac{1}{N} \fs
+\frac{1}{N} \ft
-\frac{(N^2+1)}{N} \fu
\Bigg\} ,
\end{equation}
and 
\begin{equation}
\overline{{\cal IR}}_{nt,C} =\frac{e^{\ep \gamma}}{\Gamma(1-\ep)}
\left( 
\frac{1}{\ep^2} +\frac{3}{2 \ep}
\right) 
 \Bigg\{
\frac{N^2-1}{N} \fs
-\frac{1}{N} \ft
+\frac{1}{N} \fu
\Bigg\}.
\end{equation}
As before, we can rewrite
Eq.~(\ref{eq:like}) directly in terms of $\overline{{\cal IR}}_{t, C}$ and
$\overline{{\cal IR}}_{nt, C}$ rather than $\overline{{\cal IR}}_{t}$ and
$\overline{{\cal IR}}_{t, C}$ provided the finite difference 
is evaluated through to ${\cal O}(\ep^2)$.

\section{Conclusions}
\label{sec:conclusions}

In this paper we discussed the ${\cal O}(\alpha_s^4)$ virtual corrections to
like and unlike massless quark-quark scattering formed by the interference  of
one-loop amplitudes with one-loop amplitudes. Our main results are
Eqs.~(\ref{eq:unlike}) and~(\ref{eq:like}) where we provided analytic formulae
for the \MSbar-renormalised amplitudes in terms of two one-loop master
integrals, the box graph in $6-2\ep$ dimensions and the bubble diagram in $4-2
\ep$ dimensions.  Expressions for the crossed and time reversed processes can
be simply produced by inserting the expansions of the master integrals in the
appropriate kinematic region. Together with the interference of two-loop
diagrams with tree graphs computed in~\cite{qqQQ,qqqq}, the one-loop square
contributions given in Eqs.~(\ref{eq:unlike}) and~(\ref{eq:like}) complete  the
set of $2 \to 2$ virtual corrections for massless quark-quark scattering at
NNLO.

\section*{Acknowledgements}

C.A. acknowledges the financial support of the Greek Government and M.E.T.
acknowledges financial support from CONACyT and the CVCP.  We gratefully 
acknowledge the support of the British Council and German Academic Exchange
Service under ARC project 1050.  This work was supported in part by the EU
Fourth Framework Programme `Training and Mobility of Researchers', Network
`Quantum Chromodynamics and the Deep Structure of Elementary Particles',
contract FMRX-CT98-0194 (DG-12-MIHT).

\end{document}